\documentclass[twoside,leqno,twocolumn]{article}

\usepackage[letterpaper]{geometry}
\usepackage{ltexpprt}
\usepackage{hyperref}
\usepackage{enumitem}
\usepackage{tablefootnote}
\usepackage{amsmath}
\usepackage{amssymb}
\usepackage{xcolor}
\usepackage{booktabs}
\usepackage[flushleft]{threeparttable}
\usepackage{graphicx}
\usepackage{chngcntr}
\usepackage{caption}
\usepackage[multiple]{footmisc}

\usepackage{algorithm}
\usepackage{algorithmic}
\usepackage{newfloat}
\usepackage{listings}

\begin{document}

\newcommand\Alpha{\mathrm{A}}
\newcommand\Beta{\mathrm{B}}

\newcommand\relatedversion{}
\renewcommand\relatedversion{\thanks{The full version of the paper can be accessed at \protect\url{https://arxiv.org/abs/1902.09310}}} 


\title{\Large SAICL: Student Modelling with Interaction-level Auxiliary Contrastive Tasks for Knowledge Tracing and Dropout Prediction}
\author{Jungbae Park\footnotemark[1]\footnotemark[2], Jinyoung Kim\footnote{RIIID AI Research}, Soonwoo Kwon\footnotemark[1], Sang Wan Lee\footnote{KAIST}}

\date{}

\maketitle


\fancyfoot[R]{\scriptsize{preprint under review.}}





\begin{abstract} \small\baselineskip=9pt

Knowledge tracing and dropout prediction are crucial for online education to estimate students' knowledge states or to prevent dropout rates. While traditional systems interacting with students suffered from data sparsity and overfitting, recent sample-level contrastive learning helps to alleviate this issue. One major limitation of sample-level approaches is that they regard students' behavior interaction sequences as a bundle, so they often fail to encode temporal contexts and track their dynamic changes, making it hard to find optimal representations for knowledge tracing and dropout prediction.
To apply temporal context within the sequence, this study introduces a novel student modeling framework, SAICL: \textbf{s}tudent modeling with \textbf{a}uxiliary \textbf{i}nteraction-level \textbf{c}ontrastive \textbf{l}earning. In detail, SAICL can utilize both proposed self-supervised/supervised interaction-level contrastive objectives: MilCPC (\textbf{M}ulti-\textbf{I}nteraction-\textbf{L}evel \textbf{C}ontrastive \textbf{P}redictive \textbf{C}oding) and SupCPC (\textbf{Sup}ervised \textbf{C}ontrastive \textbf{P}redictive \textbf{C}oding). While previous sample-level contrastive methods for student modeling are highly dependent on data augmentation methods, the SAICL is free of data augmentation while showing better performance in both self-supervised and supervised settings. By combining cross-entropy with contrastive objectives, the proposed SAICL achieved comparable knowledge tracing and dropout prediction performance with other state-of-art models without compromising inference costs.
\end{abstract}

\section{Introduction.}

\begin{figure*}[t]
\centering
\includegraphics[width=0.99\textwidth]{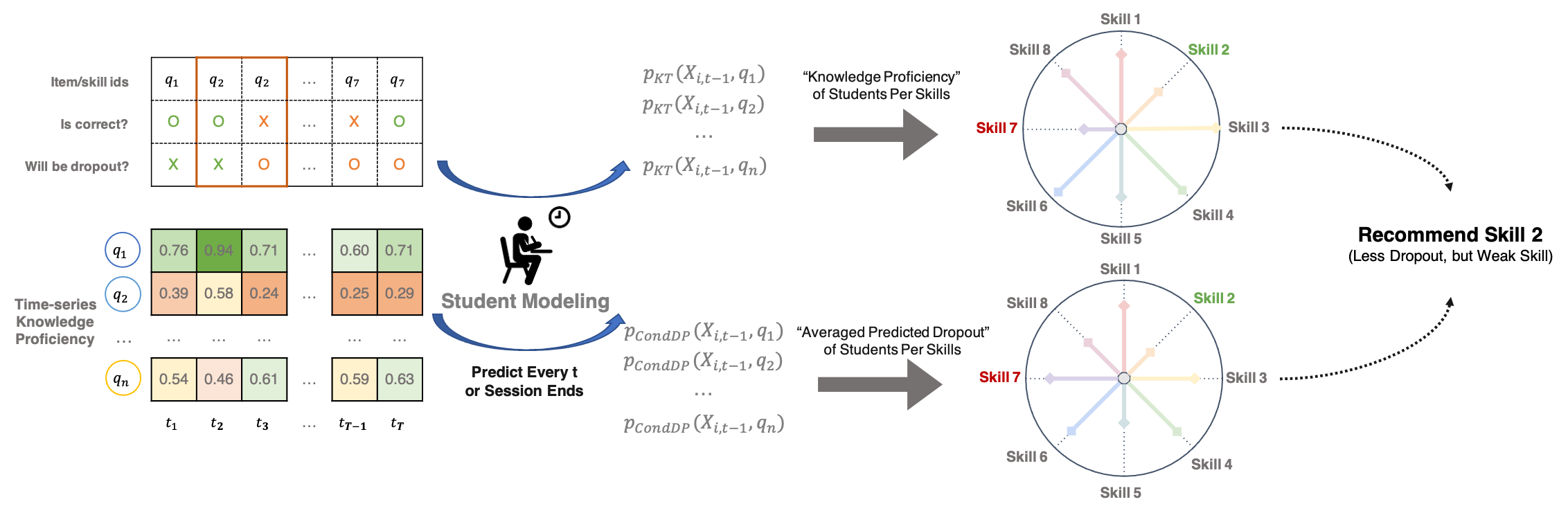}
\caption{An illustration of how student modeling is used for ITS. The left of the figure shows the student's historical interactions. Given student interaction, for every timestamp, student models such as knowledge tracing and dropout prediction are used to trace student knowledge states or dropout probabilities. After aggregating the predicted probability for each item (the right of the figure, see radar diagrams), the system can decide which contents or skills are appropriate for students. Since the procedure typically requires prediction for conditioned all skills or questions for \textbf{every timestamp}, it is essential to \textbf{predict all cases efficiently} and \textbf{consider temporal contexts} dynamically. Meanwhile, the previous global-aggregated sample-level CL approaches cannot distinguish each interaction even with temporal contexts, as shown in the orange box in the left figure.}

\label{fig:studentmodling}
\end{figure*}

Online education platforms, e.g., MOOC and intelligent tutoring systems (ITS), have received rapt attention due to the global pandemic. These techniques require collecting students' behavior data, enabling large-scale personalized learning. An adaptive instructional system such as an ITS manages teaching and learning processes by observing each student's behavior changes to make an inference on needs, preferences, or learning styles appropriate for each student. Student models, represented as knowledge tracing and dropout prediction, are essential for scalable personalized education.

However, mining students' behaviors suffer from sparsity, making it difficult to recognize patterns reflecting student education status. Contrastive learning (CL) is one remedy for this issue \cite{chopra2005origin_contrastive}. CL methods generally learn a representation by pulling an anchor and a positive sample together in the embedding space while pushing apart the anchor from many negative samples. Despite the high computational cost of CL due to its matrix calculation of similarity among embeddings, CL is widely used because the inference stage or downstream tasks do not require the computing of operation for CL.

For existing previous methods for sequential data, a naive sample-level CL approach like SimCLR \cite{chen2020simclr} can be found in \cite{xie2020contrast_seq_rec}. Local interaction contexts are aggregated into global to utilize sample-level CL. Recently proposed CL4KT \cite{lee2022cl4kt} is also based on this schema and does not consider the temporal contexts. However, for systems where temporal contexts are essential, such as education (see Fig. \ref{fig:studentmodling}), the former method might not be sufficient to learn local representations since the global aggregation bunches representations within sequences (see Fig. \ref{fig:tsne_plot}). This issue can be crucial because local interval intervention from the tutoring system is common, and students' knowledge states can be changed during the study.  

To address the issues, we propose the novel interaction-level CL for student modeling on both self-supervised and supervised setups. For the self-supervised configuration, each interaction representation combination within a single user is pulled together, and the interactions from other users are repelled. This helps the model distinguish user behavior, enabling finding the local optima across sessions. On the other hand, the supervised setup tries to catch the dynamic change by leveraging label information as \cite{khosla2020supcontrast}, helping the model understand interval intervention from the tutoring system. Then, based on suggested interaction-level CL approaches, we introduce a novel educational student modeling framework for learning temporal representations, SAICL. Next, we compare our methods with baselines on both knowledge tracing (KT), dropout prediction (DP), and conditional dropout prediction (CondDP). The proposed frameworks perform better than baselines and global-aggregated CL without adding extra features or data augmentations. Lastly, we present quantitative results to show the robustness of our methods.


\section{Problem Formulation and Related Works}

This paper focuses on knowledge tracing, dropout prediction, and conditional dropout prediction among several sequential student modeling approaches. First, we define knowledge tracing and dropout prediction, the problem the proposed models aim to address. Second, we discuss previous studies on CL methods. 

\subsection{Student Modeling}
\subsubsection{Knowledge Tracing}
Knowledge tracing (KT) refers to inferring whether students will answer the question correctly or not, based on their estimated knowledge state 
from past interactions. For the user $i \in I$, let student interaction as $x_{i, t} = (q_{i, t}, a_{i, t}) \in X$, sets of item (e.g., questions, skill) information ($q_{i, t}$) and user response information ($a_{i, t}$), where $t \in T_i$ is relative activity time of the user $i$. The response ($a_{i, t}$) accompany correctness label information ($y_{i, t}  \in \{0, 1\}$) and other auxiliary features such as timestamp or response time. Then knowledge tracing is specified as a general conditional classification problem, aiming to estimate the following conditional probability:
\begin{equation}
\begin{aligned}
    \mathrm{P}(y_{i, t} = 1|x_{i, 1}, ..., x_{i, t-1}, q_{i, t}).
\end{aligned}
\end{equation}
While other features like elapsed time, course information, or user metadata can be used, in this study, for simplicity, we only use question or skill ids and correctness like conventional studies \cite{piech2015dkt, pandey2019sakt}.

\begin{figure*}[!ht]
    \centering
    \includegraphics[width=0.95\textwidth]{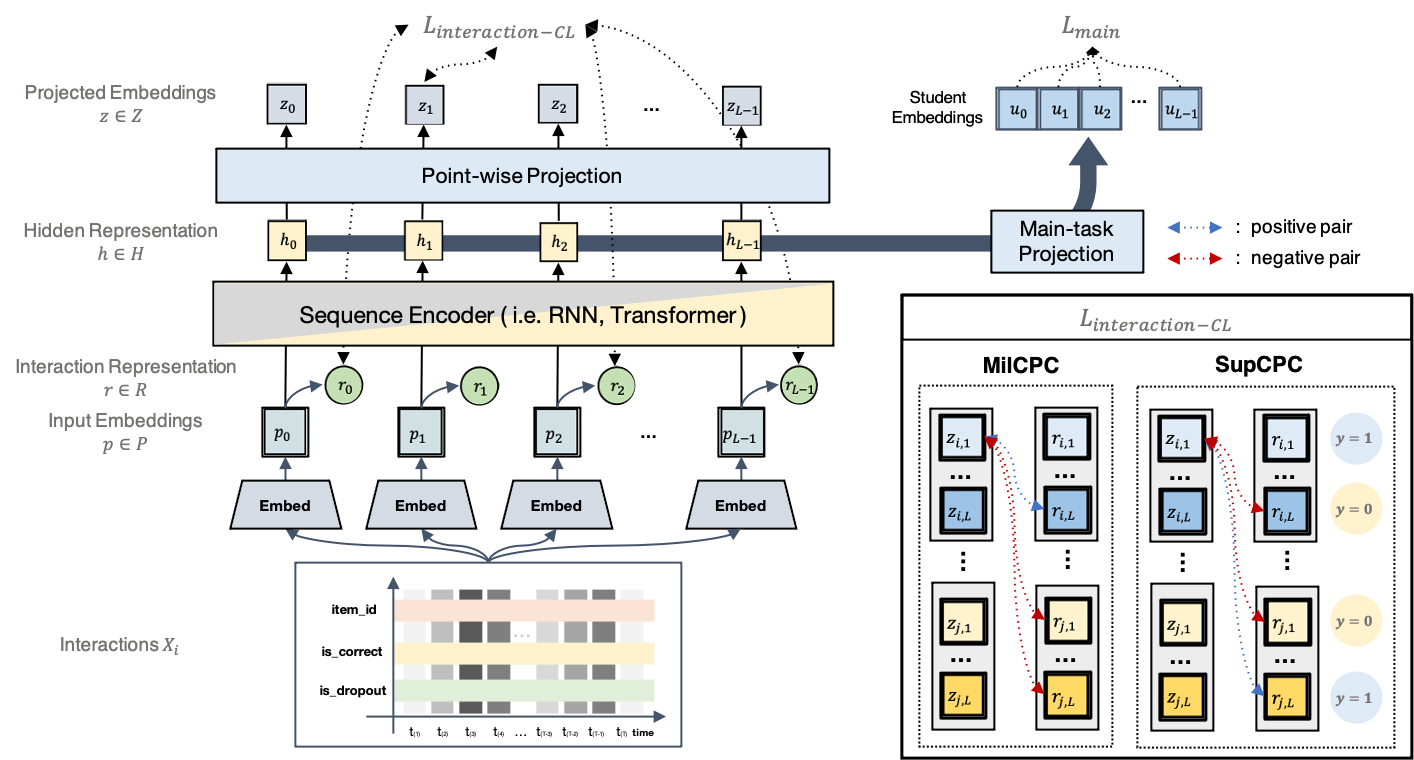}
    \caption{Proposed frameworks (left) with the illustration of interaction-level contrastive methods (right).}
    \label{fig:pcm4sm}
\end{figure*}

\subsubsection{Dropout Prediction}
Dropout prediction (DP) refers to a sequential classification problem, inferring whether a student will drop out in the future based on students' logs. In this work, we consider two dropout prediction tasks: 1) DP for the MOOC, a general sequence classification problem, and 2) conditional DP for the mobile educational platform, aiming to identify contents triggering students' dropout. 

The general problem formulation for dropout prediction on the MOOC can be found in \cite{feng2019understanding}. Given the user's learning interaction activity on any course ($x_{i, t}$) in the previous period ($t_h$, implying $t \leq t_h$), dropout prediction aims to predict whether the student will drop out from the course after the prediction period ($t_p$):

\begin{equation}
    \mathrm{P}(y_{i, t_h + t_p} = 1|x_{i, 1}, ..., x_{i, t_h}),
    \label{eq:dropout_form}
\end{equation}
where $y_{i, t_h + t_p} \in \{0, 1\}$ is a binary index showing whether student drops out of the course at the $t_h + t_p$.

On the other hand, conditional dropout prediction for mobile educational platforms can be found in \cite{lee2020sessiondropout-das}. While earlier works focused on predicting session dropout, it has limited applicability for the following reasons. First, users can decide dropout from the applications even before the model notices the dropout event. Second, educational content recommendation systems on the market cannot afford to change curricula suddenly. Lastly, session dropout data often suffers from a label imbalance problem. To address these issues, we generalize the conditional dropout prediction problem to predict the dropout probability in the earlier stage and to make pedagogical recommendation agents easier to change the curriculum with comparably balanced dropout labels. Consider the user's learning interaction activity on the any course ($x_{i, t}$) in history period ($t_h$, implying $t \leq t_h$). Similar KT, the student interaction is composed of sets of item (e.g. questions, skill) and response, $x_{i, t} = (q_{i, t}, a_{i, t}) \in X$. Then the item-conditioned dropout prediction can be defined by modifying from the Eq \ref{eq:dropout_form}:

\begin{equation}
    \mathrm{P}(y_{i, t_h + t_p} = 1|x_{i, 1}, ..., x_{i, t_h}, q_{i, next}),
    \label{eq:cond_dropout_form}
\end{equation}
where $q_{i, next}$ is the next item information after $t_h$. Note that while \cite{lee2020sessiondropout-das} defines the conditional dropout relatively for the one after the previous interactions, Eq \ref{eq:cond_dropout_form}. infers the probability of dropout of the user after some absolute times ($t_p$) from the last interacted moment ($t_h$) on user activity history.

\subsection{Contrastive Learning for Student Modeling}

Since students' historical data consists of temporal interaction sequences, sample-level contrastive learning like SimCLR \cite{chen2020simclr} cannot be directly applied. To resolve issues, inspired from CL4SRec \cite{xie2020contrast_seq_rec}, CL4KT \cite{lee2022cl4kt} aggregates all local interaction representations from each interaction into global and applying sample-level contrastive methods like static domains. 

The objective function of this case can be defined as follows. Let $\tilde{x}_{i, t, (\cdot)}\in \tilde{X_i}$ be the arbitrary augmented interactions from $X_i$. If the augmented samples are from the same user, this pair is marked as positive, and the other pairs are negative. Let denote, within a jointly augmented batch, $pos(i)$ is the index of the other augmented (positive) sample anchoring $i\in I$. If $z_i$ is embedding vectors from $\tilde{X_i}$ through encoder and projections, meaning $z_i = Proj_{out}(SeqEnc([\tilde{x}_{i, 1, (\cdot)}, ..., \tilde{x}_{i, t, (\cdot)}]))$, then the contrastive object can be defined as follow:

\begin{equation}
\begin{split}
  L_{Concat-InfoNCE} = 
  & - \sum_{i}\log \dfrac{\exp(z_i\cdot z_{pos(i)}/\tau)}{\sum_{\gamma\in \Gamma(i)} exp(z_i \cdot z_\gamma / \tau)},
\label{eq:simclr}
\end{split}
\end{equation}
where $\Gamma(i) \equiv I\setminus \{i\}$ and $\tau \in R^+$ is the scalar temperature hyperparameter.

These global-aggregated approaches relieve data sparsity on user interaction data; however, these sample-level methods have two shortcomings. First, for the cases where temporal contexts are important, like education, this former method might be insufficient to learn optimal representations of the sequences. Second, these methods depend highly on data augmentation methods requiring many hyperparameter tunings. On the other hands, based on graph contrastive learning \cite{you2020graph_contrastive}, Bi-CLKT \cite{song2022bi-clkt} proposes graph-based CL. However, as GKT \cite{nakagawa2019gkt}, constructing graph structures are computationally expensive and too slow to apply to large-scale educational platforms like MOOC, which requires inferring the correct probabilities according to the conditions of all items or skills (see Fig. \ref{fig:studentmodling}).

\section{Proposed Methods}


In this section, we introduce our proposed framework, SAICL (see Fig. \ref{fig:pcm4sm}) with suggested interaction-level CL.

\subsection{Model Architecture}

\subsubsection{Input Embedding for Interaction, $Enc_{in}(\cdot)$}
The students' historical interaction data consists of multi-column inputs. Each column of inputs $x$ can be categorized into:

\begin{itemize}
    \item Categorical, Position Feature ($F_{cb}$): e.g. item (question) ids, correctness, dropouts, position ...
    \item Continual Real Feature($F_{cr}$): e.g. elapsed-time, lag-time, ...
\end{itemize}
We assume that all features of the user interaction data are sampled or preprocessed by adding pad values to have the same length. The input embedding of the proposed backbone is defined as follows: 
\begin{equation}
\begin{split}
    p_{i, t} &= Enc_{in}(x_{i, t}) \\
    &=\sum_{cb\in F_{cb}} (W_{in}^{cb} \cdot OneHot(x_{i, t}^{cb})) + \sum_{cr\in F_{cr}} (W_{in}^{cr} \cdot x_{i, t}^{cr}),
\end{split}
\end{equation}
where $x_{i, t} \in x_i$ is a interaction of $i$-th student at position $t$ and $W_{in}^{cb}$, $W_{in}^{cr}$ are trainable parameters.

\subsubsection{Backbone Sequential Encoder, $SeqEnc(\cdot)$} According to \cite{oord2018cpc}, any autoregressive model for interaction-level contrastive learning can be used for temporal context embeddings. In detail, for each domain, we choose the backbone as follows.
\\
$SeqEnc_{KT}(\cdot):$ While several methods are proposed for KT, long-short term memory (LSTM)-based DKT model \cite{hochreiter1997lstm, piech2015dkt} is still comparable with other baselines. Since DKT is faster than other recent works, we choose a simple LSTM sequence encoder for this study.
\\
$SeqEnc_{DP}(\cdot):$ Since context-aware feature interaction network (CFIN) \cite{feng2019understanding} previously achieved the best performance on DP, CFIN itself does not have temporal aggregation except attentional weighted sum. To provoke the model to understand the temporal context better, we propose SAEDP (self-attentive encoder for dropout prediction) backbone, utilizing a transformer encoder \cite{vaswani2017transformer}. For more details, please refer to the appendix.
\\
$SeqEnc_{CondDP}(\cdot):$ A simple transformer encoder with a causal mask is used for CondDP.

Commonly, from the input embeddings, $p_i$, shared $SeqEnc$ maps both augmented samples to a representation space at the hidden layer, as $h_{i, t} = SeqEnc(p_{i, 1:t})$. $h_{i, t}$ will also be used for downstream tasks after pretraining or main tasks as multi-task learning. For more details about the hyperparameters of each task, please look up the appendix.

\subsubsection{Point-wise Output Projection, $Proj_{out}(\cdot)$}

For embeddings for contrastive objective, $Proj_{out}(\cdot)$ maps the hidden representation $h_{i, t}$ into the projected output vector $z_{i, t} = Proj_{out}(h_{i, t})$. The $z_{i, t}$ is used for the training stage only and is discarded at the end of contrastive learning. 

\subsubsection{Point-wise Interaction Projection, $Proj_{inter}(\cdot)$} Like wav2vec 2.0 \cite{baevski2020wav2vec} on audio domain, for contrastive objectives, input interaction embeddings $p_{i, t}$ is forwarded into interaction projection rather than using same embeddings for target of contrastive objects like \cite{chen2020simclr}. The reference interaction representation $r_{i, t}$ = $Proj_{inter}(p_{i, t})$ will be used for target of contrastive objectives. Like source inputs for contrastive objective $z_{i, t}$, $r_{i, t}$ will be left out for main tasks.

\subsection{Auxiliary Interaction-level Contrastive Learning}

 We introduce two types of auxiliary interaction-level contrastive methods, self-supervised setting and supervised setting for student modeling. 

\subsubsection{Multi-interaction-level CPC}

While the objective function of contrastive predictive coding (CPC) \cite{oord2018cpc} can learn temporal contexts for self-supervised representation learning, it is limited to taking only one positive interaction per source. Practically, multiple positive interactions can be defined within the same interactions as \cite{qiu2021micpc}. Let $\Upsilon: I \times T \equiv \{(i, t)\mid i\in I, t \in T_i \}$, $\Gamma(i, t) \equiv \Upsilon\setminus\{(i, t) \}$, and $A(i, t) \equiv \{ (\alpha, t_m) \in \Gamma(i, t): \alpha = i \} $ be the set of indices of all positives in the multi-view batch and across sequences, anchored from $i, t$. That is taking positive interactions from the same user sequences to make the learning process consistent, but to make the interaction apart from other users' interactions. Then the loss $L_{MilCPC}$ can be defined as follow:
\small
\begin{equation} \label{eq:micpc}
\begin{split}
    &L_{MilCPC} \\
    &= \sum_{i, t}\frac{-1}{|A(i, t)|}\sum_{\alpha, t_m \in \Alpha(i, t)} \log{\frac{\exp (z_{i, t} \cdot r_{\alpha, t_m} / \tau)}{\sum_{\gamma \in \Gamma(i, t)}\exp (z_{i, t}\cdot r_\gamma / \tau)}},
\end{split}
\end{equation}
\normalsize
where $|A(i, t)|$ is cardinality of $A(i, t)$. For this case, pretraining loss objectives can be defined as follow:
\small
\begin{equation}
\begin{split}
    L = L_{CE} + \lambda_{self} L_{MilCPC},
\end{split}
\label{eq:ce_micpc}
\end{equation}
\normalsize
where $L_{CE}$ is the cross-entropy loss, commonly used for knowledge tracing and dropout prediction, and $\lambda_{self}$ is modulating hyperparameter.


\subsubsection{Supervised CPC}
To leverage label information for interaction-level CL, we extend SupContrast \cite{khosla2020supcontrast} to support temporal information, called supervised contrastive predictive coding (SupCPC). Let $\Upsilon: I \times T \equiv \{(i, t)\mid i\in I, t \in T_i \}$, $\Gamma(i, t) \equiv \Upsilon\setminus\{(i, t) \}$, and $P(i, t) \equiv \{(\alpha, \beta) \in \Gamma(i, t): y_{\alpha, \beta} = y_{i, t} \} $ be the set of indices of all positives in the mini-batch and across sequences, anchored from $i, t$. Please note that unlike $L_{MilCPC}$, the mining of positive and negative interactions is determined by labels, not predefined superpositions (i.e., from the same user). Then the SupCPC objective equation can be written as follow:
\small
\begin{equation}
\begin{split}
    & L_{SupCPC} \\ 
    & = \sum_{i, t} \frac{-1}{|P(i, t)|} \sum_{p\in P(i, t)}\log{\frac{\exp (z_{i, t} \cdot r_{p} / \tau)}{\sum_{\gamma\in \Gamma(i, t)}\exp ( z_{i, t} \cdot r_\gamma / \tau)}},
\end{split}
\label{eq:SupCPC}
\end{equation}
\normalsize
where $|P(i, t)|$ is cardinality of $P(i, t)$. Since tasks of KT and CondDP are getting conditional probability for the next item, to consider conditional items more, we propose conditional supervised contrastive predictive coding (C-SupCPC), based on conditional contrastive learning (CCL) \cite{tsai2021condcl}. If we set conditional input as item, then positive samples can be defined as $P_c(i, t) \equiv \{(\alpha, \beta) \in \Gamma(i, t): y_{\alpha, \beta} = y_{i, t} \wedge q_{\alpha, \beta} = q_{i, t} \} $. The Eq. \ref{eq:SupCPC} can be rewritten as follow:
\small
\begin{equation}
\begin{split}
    & L_{C-SupCPC} \\ 
    & = \sum_{i, t} \frac{-1}{|P_c(i, t)|} \sum_{p\in P_c(i, t)}\log{\frac{\exp (z_{i, t} \cdot r_{p} / \tau)}{\sum_{\gamma\in \Gamma(i, t)}\exp ( z_{i, t} \cdot r_\gamma / \tau)}}.
\end{split}
\label{eq:C-SupCPC}
\end{equation}
\normalsize
By combining both unconditioned and conditioned cases, the pretraining objectives can be defined as: 
\small
\begin{equation}
\begin{split}
    L = L_{CE} + \lambda_{sup} L_{SupCPC} + \lambda_{sup} L_{C-SupCPC},
\end{split}
\label{eq:ce_supcpc}
\end{equation}
\normalsize
where $\lambda_{sup}$ is a modulating hyperparameter.

\section{Experiments}

We verify the performance of our frameworks with real-world data with a series of experiments. First, we compare how sample-level and interaction-level CL objectives learn the representations by training separately with t-SNE plots. Second, we evaluate our methods with other state-of-models for KT, DP, and CondDP. Next, we compare proposed interaction-level CL objectives with different CL baselines with the fixed encoder setups. Lastly, we analyze the difference in the performance as the hyperparameter changes, including data augmentations as ablation tests.

\subsection{Datasets}

To evaluate proposed methods, we use following open benchmark datasets for KT: ASSISTments(2009, 2015)\footnote{https://sites.google.com/site/assistmentsdata} \cite{feng2009assistments} and STATICS2011\footnote{https://pslcdatashop.web.cmu.edu/DatasetInfo?datasetId=507}. We removed the user whose number of interactions is lower than 5 and used skill tags for conditional input as the convention. For tests, we only use columns of the student, question id(or skill tags), and correctness information. Afterward, we randomly split user data in the ratio of 72:8:20 for training/validation/testing. For the evaluation, we evaluated all interactions individually, one by one, including the starting point of the sequence.

We used two dropout benchmark datasets for DP on MOOC: KDDCup2015\footnote{https://www.biendata.xyz/competition/kddcup2015/} and XuetangX\footnote{http://moocdata.cn/data/user-activity}. 
The history period ($t_h$) and the prediction period ($t_p$) are set as $t_h=30\ days, t_p=10\ days$ for KDDCup2015 and $t_h=35\ days, t_p=10\ days$ for XuetangX as \cite{feng2019understanding}.

For CondDP, we use EdNet\footnote{https://github.com/riiid/ednet} with preprocessing dropout labels, conditioning $t_h=30\ days, t_p=7\ days$ for evaluations. Processed EdNet-DP will be available with implemented source code. We removed the user whose number of interactions is less than 50 or whose activity date is less than 7 days.

For detailed statistics, please refer to the appendix.

\subsection{Experiment Setups}

We suggest a simple training pipeline as follows. First, all interactions of user interactions for a batch are retrieved. To construct ‘length-invariant’ sequences for the training loop, we randomly extract the consequent interaction size of $L$, the same as the model sequence size on a batch. Then we apply min-max normalization for continuous real features in advance to prevent changes in padding values. Then, if the interaction size is smaller than the model sequence size ($T_i \leq L$), we add padding values to the end of the sequence to match the sequence length. When finalizing the matching length of whole sequences within the batch, we apply additional feature engineering logic like inserting start-tokens. Finally, preprocessed interaction features are ready to be input into the models.

For optimizations, we use RAdam \cite{liu2019radam} optimizer for all experiments, which is rectified version of Adam \cite{kingma2014adam} to reduce the initial convergence problem by reducing the initial variance of gradients. This also enables reliable training (e.g., less sensitivity to the choice of the learning rate or warm-up scheduling).

\subsection{Representation t-SNE plots}

\begin{figure*}[!hbt]
\centering
\includegraphics[width=0.95\textwidth]{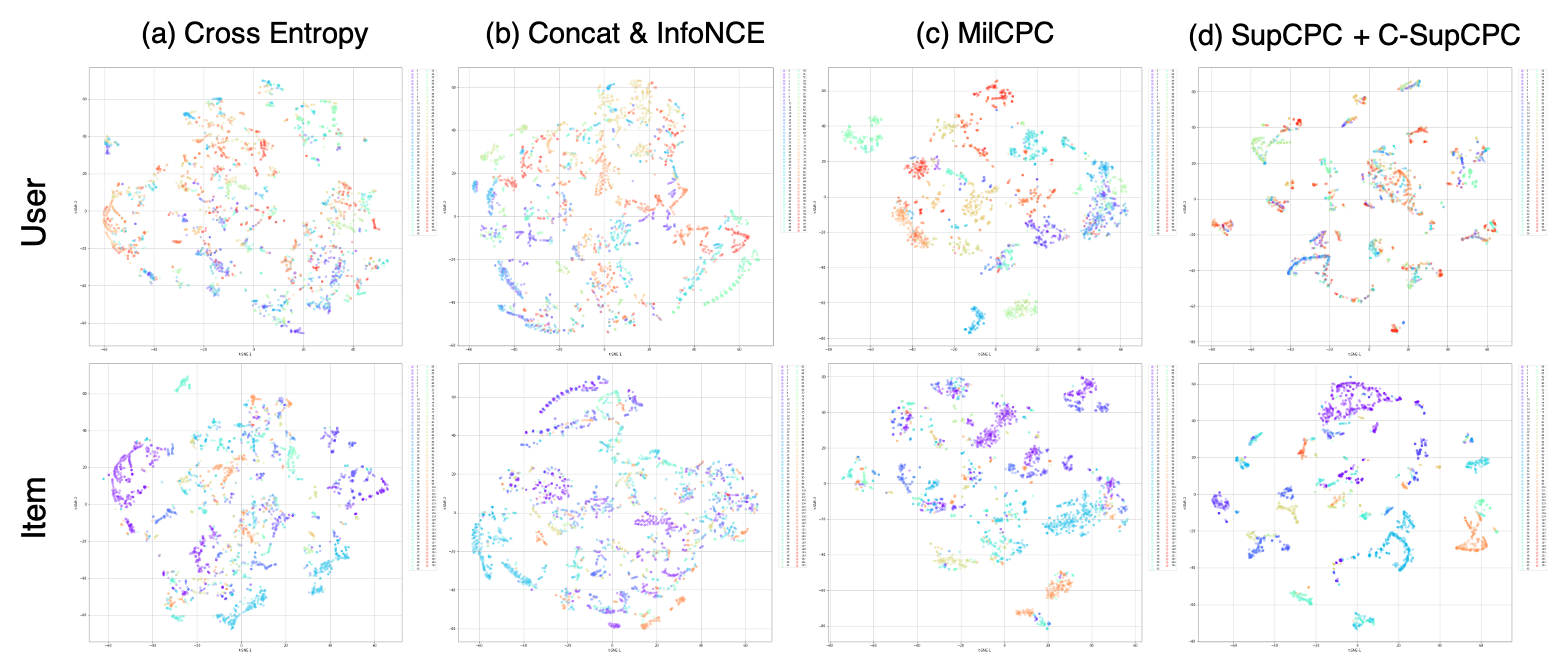}
\caption{t-SNE plots for cross-entropy and contrastive methods, introduced in this study on KT tasks trained on ASSISTments2015 data. While the cross-entropy and concatenated \& InfoNCE fail to split representations of users, MilCPC obtains better representation regarding students' personalization. However, for SupCPC, it does not seem to discern each student's representation because it does not pull or repel for user-level indices. Instead, it is generally better to distinguish conditioned items.}
\label{fig:tsne_plot}
\end{figure*}

To analyze how each CL method affects student modeling, we train each CL objective and baseline cross-entropy separately, and plot hidden representations with t-SNE \cite{van2008tsne} (see Fig. \ref{fig:tsne_plot}). We randomly sampled 100 users and used the most recent interactions from the test dataset but filtered padded masks. While the baseline cross-entropy or the sample-level CL (Concat \& InfoNCE) failed to optimally separate users or items, MilCPC helps to learn the distinguishments of each student, which can be essential for personalized education. On the other hand, SupCPC seemingly helps learn local conditioned question inputs.

\begin{table*}[htb!]
\small

  \centering
  \caption{
    Benchmark KT with projection performance to get all items
  }
  \resizebox{\textwidth}{!}{
 \begin{tabular}{lcccccccc}

\toprule
 &  & \multicolumn{2}{c}{ASSISTments 2009} & \multicolumn{2}{c}{ASSISTments 2015} & \multicolumn{2}{c}{STATICS 11} \\
\midrule
Model & Cost To Predict All Items & ACC (\%) & AUC (\%) & ACC (\%) & AUC (\%) & ACC (\%) & AUC (\%) \\ 
\midrule
SAKT & $\mathcal{O}(Q\times (LD^2 + D))$ & $74.23_{\pm 0.01}$ & $78.55_{\pm 0.01}$ & $72.51_{\pm 0.01}$ & $69.66_{\pm 0.01}$ & $79.56_{\pm 0.23}$ & $79.48_{\pm 0.13}$ \\ 
SAINT$^{\dagger}$ & $\mathcal{O}(Q\times (LD^2 + D))$ & $74.46_{\pm 0.01}$ & $78.23_{\pm 0.01}$ & $71.76_{\pm 0.01}$ & $67.52_{\pm 0.01}$ & $78.40_{\pm 0.55}$ & $77.75_{\pm 0.15}$ \\ 
AKT & $\mathcal{O}(Q\times (LD^2 + D))$ & $76.02_{\pm 0.10}$ & $81.05_{\pm 0.08}$ & $\mathbf{72.97_{\pm 0.01}}$ & $71.25_{\pm 0.01}$ & $\mathbf{80.64_{\pm 0.07}}$ & $82.13_{\pm 0.09}$ \\ 
CL4KT & $\mathcal{O}(Q\times (LD^2 + D))$ & $75.90_{\pm 0.28}$ & $80.89_{\pm 0.26}$ & $72.95_{\pm 0.04}$ & $71.22_{\pm 0.15}$ & $80.12_{\pm 0.17}$ & $80.78_{\pm 0.37}$\\ 
\midrule
DKT (Backbone) & $\mathbf{\mathcal{O}(L^2D + DQ)}$ & $75.53_{\pm 0.03}$ & $80.54_{\pm 0.16}$ & $72.64_{\pm 0.01}$ & $70.65_{\pm 0.01}$ & $80.02_{\pm 0.11}$ & $81.56_{\pm 0.11}$  \\
SAICL$_{self}$ & $\mathbf{\mathcal{O}(L^2D + DQ)}$ & $\mathbf{76.03_{\pm 0.03}}$  & $\mathbf{81.15_{\pm 0.06}}$ & $\mathbf{72.91_{\pm 0.04}}$ & $\mathbf{71.36_{\pm 0.12}}$ & $\mathbf{80.31_{\pm 0.17}}$ & $\mathbf{82.12_{\pm 0.15}}$ \\ 
SAICL$_{sup}$ & $\mathbf{\mathcal{O}(L^2D + DQ)}$ & $\mathbf{75.95_{\pm 0.19}}$  & $\mathbf{81.16_{\pm 0.09}}$ & $\mathbf{72.93_{\pm 0.04}}$ & $\mathbf{71.31_{\pm 0.13}}$ & $\mathbf{80.35_{\pm 0.08}}$ & $\mathbf{82.15_{\pm 0.03}}$ \\

\bottomrule
\end{tabular}%
}
\label{table:result_kt}
\normalsize
\end{table*}

\begin{table}[!h]
\small
\centering
\begin{threeparttable}
  \caption{AUC (\%) performance comparisons for dropout prediction on MOOC.}
  \begin{tabular}{lcc}
    \toprule
    Model & XuetangX & KDD15 \\ \midrule
    ConRecNet$^{**}$ & - & 87.42 \\
    CFIN$^{**}$ & 86.40 & 90.07 \\
    CFIN-ensemble$^{**}$ & \underline{86.71} & \underline{90.93} \\ \midrule
    Backbone (SAEDP) & $\mathbf{88.72_{\pm 1.84}}$ & $\mathbf{91.31_{\pm 0.62}}$ \\
    SAICL$_{self, multitask}$ & $\mathbf{89.61_{\pm 0.79}}$ & $\mathbf{91.81_{\pm 0.12}}$ \\
    SAICL$_{sup, multitask}$ & $\mathbf{90.11_{\pm 0.22}}$ & $\mathbf{91.69_{\pm 0.40}}$ \\
    \bottomrule
    \end{tabular}
    \begin{tablenotes}[flushleft]
    \footnotesize
    \item[$**$]{ means the result is from the original paper.}
    \end{tablenotes}
    \label{table:result_dp}
\end{threeparttable}
\normalsize
\end{table}

\begin{table}[!h]
\small
  \caption{AUC (\%) performance comparison for conditioned dropout prediction for EdNet-DP.}
  \centering
    \begin{tabular}{lc}
        \toprule
        Model & EdNet-DP \\
        \midrule 
        LSTM-Based & $71.89_{\pm 2.96}$ \\
        DAS & $77.18_{\pm 0.54}$ \\
        \midrule
        Backbone (Transformer$_{dec}$) & $77.04_{\pm 0.10}$ \\
        SAICL$_{self, finetune}$ & $\mathbf{80.67_{\pm 0.39}}$ \\
        SAICL$_{sup, finetune}$ & $\mathbf{79.83_{\pm 0.53}}$ \\ 
        \bottomrule
    \end{tabular}
\label{table:result_cond_dp}
\normalsize
\end{table}

\begin{table*}[!htb]
\small
\centering
\begin{threeparttable}
  \caption{AUC (\%) performance comparisons between the best performance of each contrastive method.}
    \begin{tabular}{lcccccc}
    \toprule
     & & & \multicolumn{2}{c}{KT} & DP & CondDP \\
    \midrule
    Methods & Category$^\star$ & Data Aug. & AS09 &  AS15 &  KDD15 & EdNet-DP \\ 
    \midrule 
    Backbone & - &  & $80.54_{\pm 0.16}$ & $70.65_{\pm 0.01}$ & $91.31_{\pm 0.62}$ & $77.04_{\pm 0.10}$  \\
    \midrule
    + Concat Contrast & sample, ss & \checkmark & $81.01_{\pm 0.12}$ & $71.21_{\pm 0.17}$ & $90.65_{\pm 0.62}$ & $79.79_{\pm 0.15}$  \\
    + Concat SupContrast & sample, s & \checkmark  & - & - & $90.86_{\pm 0.81}$ & -\\
    \midrule
    + MilCPC & interaction, ss &  & $\mathbf{81.15_{\pm 0.06}}$ & $\mathbf{71.36_{\pm 0.12}}$ & $\mathbf{91.81_{\pm 0.12}}$ & $\mathbf{80.67_{\pm 0.39}}$  \\ 
    + SupCPC & interaction, s & & $\mathbf{81.12_{\pm 0.02}}$ & $\mathbf{71.31_{\pm 0.13}}$ & $\mathbf{91.69_{\pm 0.40}}$ & $\mathbf{79.83_{\pm 0.53}}$ \\
    \bottomrule
    \end{tabular}
    \begin{tablenotes}[flushleft]
        \item{$*$}: "s" means supervised CL and "ss" means self-supervised CL.
    \end{tablenotes}
\label{table:comparison_contrast}
\end{threeparttable}
\normalsize
\end{table*}

\subsection{Comparisons with State-of-Arts}

For KT and CondDP, as conventional representation learning tasks \cite{chen2020simclr, khosla2020supcontrast}, we find that finetuning after pretraining contrastive objectives with CE loss to learn the hidden representation is better. For finetuning, new point-wise MLP layers are added. 

\subsubsection{Knowledge Tracing}

We compare our methods with several baselines, DKT \cite{piech2015dkt}, SAKT \cite{pandey2019sakt}, SAINT \cite{choi2020saint}, AKT \cite{ghosh2020akt}, and CL4KT \cite{lee2022cl4kt}. We fix all model sequence sizes as 100. Note that for fairness of comparisons, we only use the “item/skill” and “correctness” information though adding additional features like time information can increase performance like \cite{shin2021saint+}. 

In comparing the baselines, we also consider computational complexity for predicting all items(skills) for KT. For the recommendation system, separating item embedding from user embedding is vital for getting all rankings of preferences of items.
Similarly, splitting the embeddings of exercises into students' embeddings is essential for a large-scale ITS system to obtain the ranking of difficulties or students' knowledge status about all skills. However, the previous transformer-based knowledge retrievers (SAKT, SAINT, AKT, CL4KT) take target conditional questions with the earlier interactions during operations of the transformer for students' embedding, so it is hard to split the calculation of student embeddings and target question embeddings. Consequently, it increases the costs of predicting all items per single student embedding.
If the model sequence size is $L$, the hidden dimension is $D$, and the number of the skills/items is $Q$, the cost to predict all items/skills is $O(Q\times (LD^2 + D))$. On the other hand, because the original DKT already splits students' embeddings with item indices into projection layers, the cost to predict all items/skills is $O(L^2D + DQ)$.
Since we aim to implement a large-scale ITS system, we use the LSTM-backbone like DKT and utilize projections as follows. For output projections of our frameworks, we use the parameter-shared projection layer \cite{kang2018sasrec} as the output projection for pretraining. On the other hand, for the projection layer in fine-tuning stage for primary tasks, we replace it with the MLP layer like the original out-projection layer of DKT.

The summarized results with baselines are in Table \ref{table:result_kt}. As shown in Table \ref{table:result_kt}, our proposed methods are comparable with other state-of-arts models without increasing inference costs.

\subsubsection{Dropout Prediction}

DP on MOOC benchmark results with baselines \cite{feng2019understanding, wang2017mooc_lstm} is in Table \ref{table:result_dp}. In choosing the out-projection layer, we use attention-based weighted sum projections like CFIN \cite{feng2019understanding}. Unlike KT, since the well-trained attention-parameter of output projection affects the performances, we find that splitting pretrain, finetuning stage is ineffective. Instead, we use our interaction-level CL to multitask, which is effective enough to achieve state-of-arts performance. While our proposed backbone (SAEDP) already outperforms the previous methods, interaction-level CL improves performance more. Note that although CFIN achieves better performance with ensemble methods, our results are obtained with the single model. This implies that it would be possible to further improve the performance by combining with ensembles.

\subsubsection{Conditional Dropout Prediction}

We compare our methods with baselines, LSTM-based \cite{wang2017mooc_lstm}, and transformer-based models (DAS, \cite{lee2020sessiondropout-das}). We use item id, correctness, elapsed time, and part id information. While the base backbone model is based on a transformer encoder with a causal mask, so it is not significantly different from DAS, our method with interaction-level CL is better than other baselines (See Table \ref{table:result_cond_dp}). Note that, like KT, it's important to consider the cost to predict all items triggering the dropout of the students to get a ranking. While DAS is also suffered from splitting student embeddings with target items, SAICL separates operations of student embedding from target items, so it is computationally much cheaper ($O(LD^2 + DQ)$) than the DAS ($O(Q\times (LD^2 + D)$).

\subsection{Comparisons with Other CL Methods}

To analyze performance differences among CL methods, we compare the proposed interaction-level CL with sample-level CL with the same backbone encoder setups. For sample-level CL, we apply data augmentations to learn multi-view of samples, as in previous studies, \cite{chen2020simclr, khosla2020supcontrast, lee2022cl4kt}. For the detailed setup for each experiment, please refer to the appendix. Please note that labels exist in interaction rather than a sample for tasks of KT and CondDP, so concatenated SupContrast can not be defined. For simplicity, we only denote our basic setting for each task: finetuning results for KT and CondDP and multi-task results for DP. The summarized results can be shown in Table \ref{table:comparison_contrast}.

As reported in \cite{lee2022cl4kt}, concatenating global temporal contexts and applying contrastive objective loss also increases overall performance. However, our interaction-level contrastive methods improve performance more, not requiring any data augmentation methods.

\subsection{Impacts of Hyperparameter $\lambda$}

\begin{figure}[!htb]
\centering
\includegraphics[width=0.99\columnwidth]{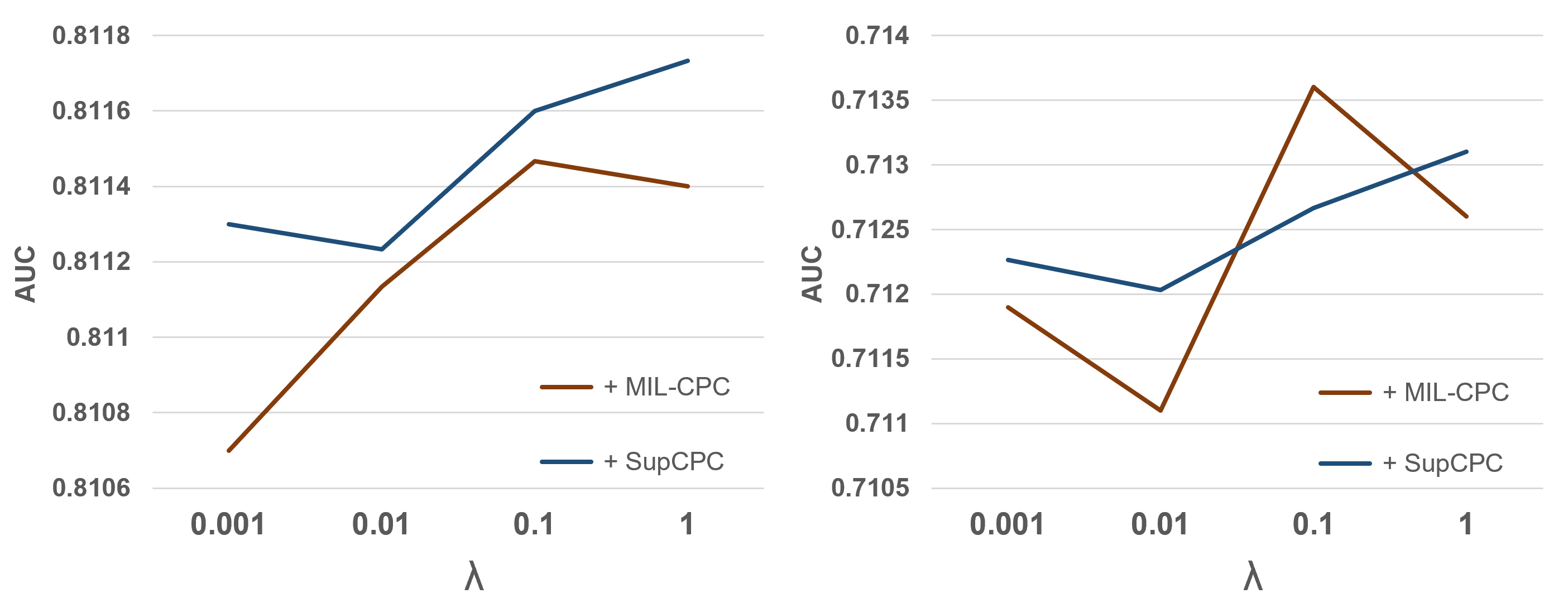}
\caption{AUC performance change with respect to $\lambda$.}
\label{fig:lambda}
\end{figure}

To compare effects of contrastive loss, we examine the influence of the CL loss by varying $\lambda$ in Eq. \ref{eq:ce_micpc} and Eq. \ref{eq:ce_supcpc} on space of $\{0.001, 0.01, 0.1, 1.0\}$. The results can be shown in Fig. \ref{fig:lambda}. We observe that for MilCPC, 0.1 is generally better than other hyperparameter setups, and for SupCPC, $\lambda$ = 1.0 makes the best performance for ASSISTments 2009, 2015 data. Though we find that the general tendency of choosing $\lambda$ follows this setup, the performance could be changed according to the characteristic of the dataset.

\subsection{Ablation Studies}

\begin{figure}[!htb]
\centering
\includegraphics[width=0.99\columnwidth]{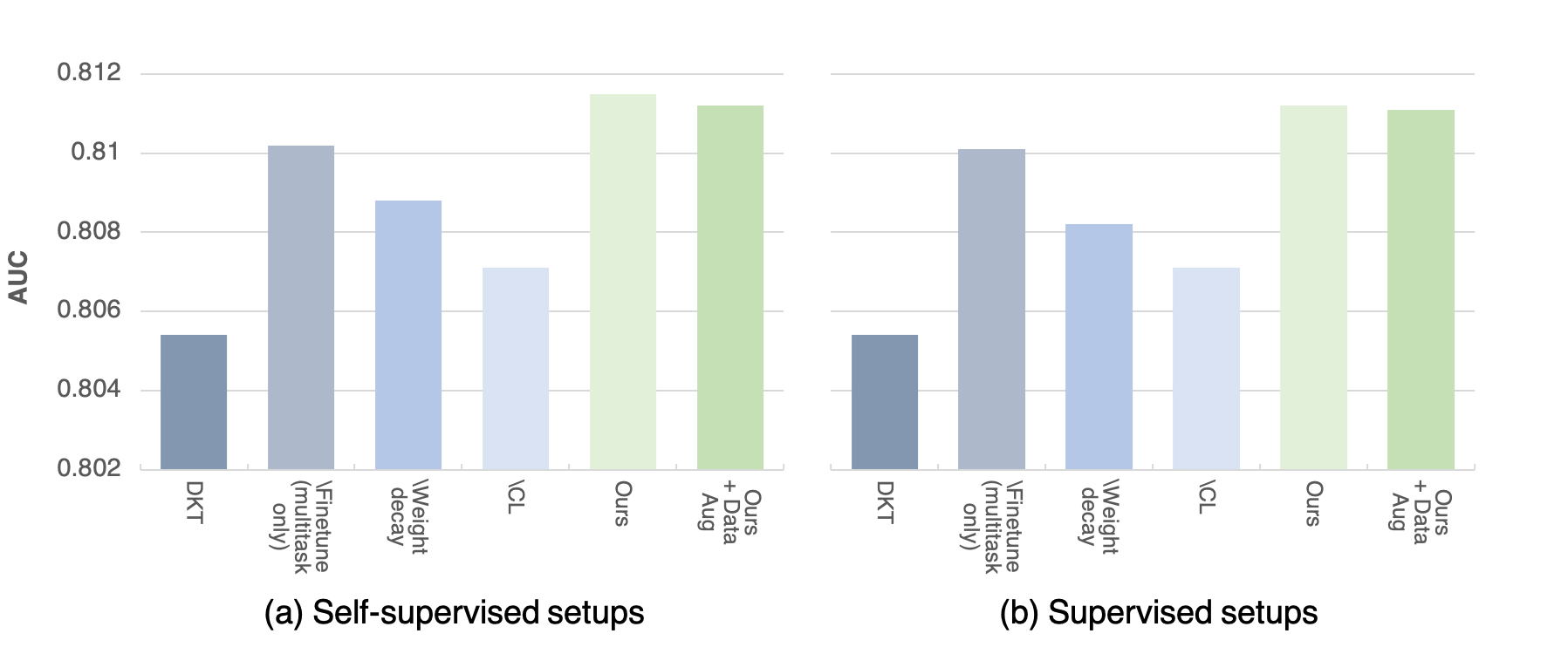}
\caption{Ablation study: performance comparison between proposed methods with other variations of KT on ASSISTments 2009 dataset.}
\label{fig:ablation}
\end{figure}

The summary of ablation studies is shown in Fig. \ref{fig:ablation}. For KT, without finetuning, the performance is slightly decreased. In addition, the performance can be reduced if there is no weight decay in the pretraining stage with CL. On the other hand, we also try to add data augmentations on interaction-level CL with the same strategy as CL4KT). However, there is no significant difference in data augmentation. It might be from the reasons that interaction-level CL tries to learn by comparing other interactions and does not essentially require different specific multi-view samples. In addition, data augmentation can increase the noise of data while the original behavior sequences of students are already noisy.

\section{Conclusions}

This study suggests a novel interaction-level CL method for student modeling, SAICL. We proposed both self-supervised (MilCPC) and supervised (SupCPC) interaction-level CL setups, which can take multiple positive samples across historical interactions, generalizing \cite{oord2018cpc}. While previous global-aggregated sample-level CL does not consider temporal contexts, interaction-level CL attempts to optimize local temporal dynamics with self-supervision or future labels. We empirically show that both methods are effective for KT, DP, and CondDP. In addition, while most previous methods do not distinguish between student embedding and item embeddings, which essentially increases the inference costs for predicting all items, SAICL achieves comparable results without compromising performances as \cite{piech2015dkt}. Also, our frameworks show effectiveness without data augmentation. It is crucial because sample-level CL should rely on data augmentation to gain multiple perspectives on the sample, requiring many hyperparameter tunings. In future works, SAICL can incorporate other sequential student modeling techniques, such as learning style modeling.

\bibliographystyle{siam}
\bibliography{bib.bib}

\newpage
\appendix
\section{Appendices}

\setcounter{figure}{0}
\setcounter{table}{0}


\subsection{Notations}

The used notations of this paper are summarized in Table \ref{table:notation}.

\begin{table}[!ht]
\small
\centering
\begin{tabular}{l|l}
    \toprule
    Notation & Definition \\
    \midrule
    $X$ & Set of students' interactions \\
    $i \in I$ & Index of an arbitrary student\\
    $t \in T_i$ & Relative activity time of the student $i$\\
    $x_{i, t} \in X_i$ & $t$-th interaction of $i$-th student ($X_i$)\\
    $\tilde{x}_{i, t, (\cdot)} \in \tilde{X_i}$ & Data augmented interactions\\
    $q \in Q$ & Item (exercise, skill-tag, item id...)\\
    $a_{i, t}$ & Response for $q_{i, t}$ (correct, elapsed time...)\\
    $(q_{i, t}, a_{i, t})$ & Item and response pair ($x_{i, t}$)\\
    $y_i$ & Label for sequence classification for $X_i$\\
    $y_{i, t}$ & Label, corresponding to $q_{i, t}$\\
    
    \bottomrule
\end{tabular}
\caption{Summary of Notations.}
\label{table:notation}
\normalsize
\end{table}


\begin{figure*}[!hbt]
\centering
\includegraphics[width=0.80\textwidth]{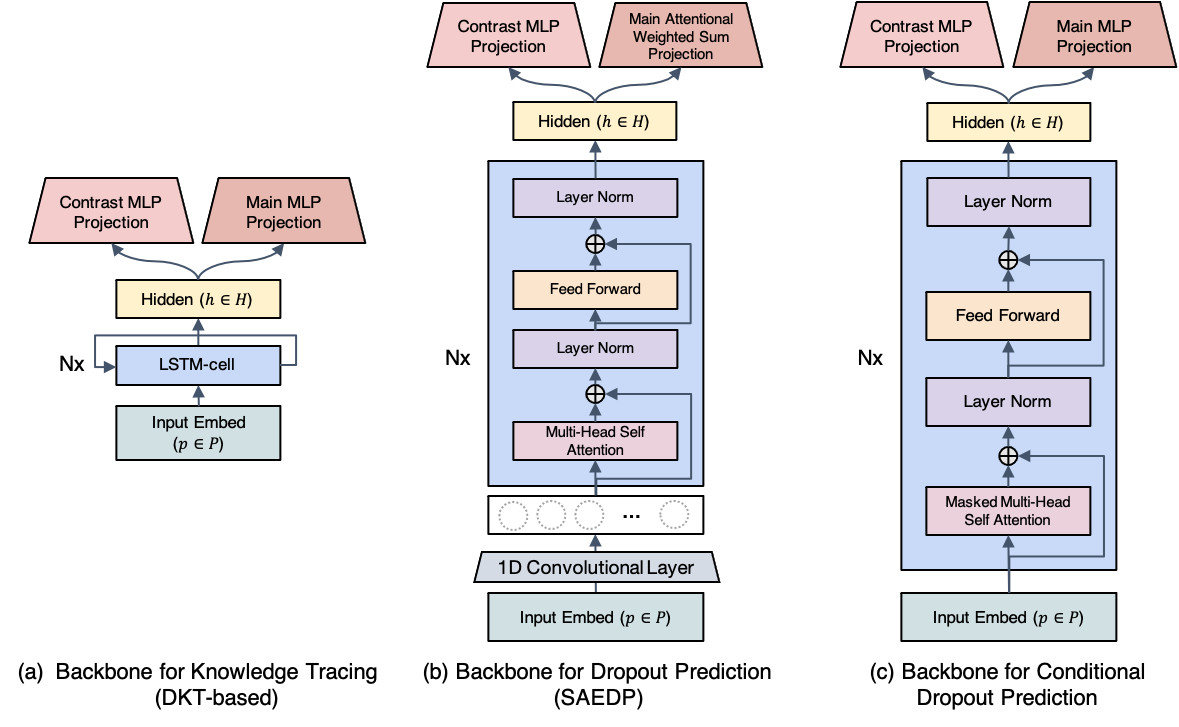}
\caption{Illustration of backbone sequence encoder models for the main experiments. Note that interaction projection, $Proj_{inter}$, is omitted from this diagram for simplicity but is connected to the input embeddings during the training stage.}
\label{fig:model_diagrams}
\end{figure*}


\subsection{Statistics of Datasets}
We follow the default setup as \cite{feng2019understanding} for dropout prediction.
The statistics of the other datasets are summarized in Table \ref{table:data_statistics}.

\begin{table}[!hbt]
  \caption{Statistics for Datasets After Preprocessing for KT and CondDP.}
  \label{table:data_statistics}
  \centering
\resizebox{0.99\columnwidth}{!}{
  \begin{tabular}{lcccc}
    \toprule
    Datasets & \# interactions &  \# users & \# items(skills) & Sparsity (\%) \\
    \midrule 
    ASSIST09 & 399,969 & 3,626 & 124 & 90.39  \\
    ASSIST15 & 699,232 & 16,858 & 100 & 92.79  \\
    STATICS11 & 188,820 & 331 & 1,212 & 53.50 \\
    \midrule
    EdNet-DP & 82,929,954 & 130,332 & 12,284 & 95.73  \\
    \bottomrule
  \end{tabular}
}
\end{table}

\subsection{Details of Backbone Sequence Encoder}

\subsubsection{LSTM-based Encoder for Knowledge Tracing}

In the experiments, we use a single layer of LSTM sequence encoder like DKT \cite{piech2015dkt} with 100 hidden sizes (see Fig. \ref{fig:model_diagrams} (a)). For fairness, we limit 100 sequence sizes for model inputs as the other baselines.

\subsubsection{Self-Attentive Encoder for Dropout Prediction}

As denoted in main page, we propose \textbf{s}elf-\textbf{a}ttentive \textbf{e}ncoder for \textbf{d}ropout \textbf{p}rediction (SAEDP) to understand the temporal context better (see Fig. \ref{fig:model_diagrams} (b)). From embeddings $p$, the 1D convolutional layers are applied. We use three continuous 1D convolutional networks having channel sizes of (32, 16, 32). Each convolutional layer has 7 filter sizes, and we apply zero padding to keep the size and position of the features. Between convolutional networks, 1d batch normalization and ReLU activation functions are applied. After then, we apply a transformer encoder to learn sequential contexts. The transformer encoder has 4 multi-heads with 128 feedforward dimensions, and a dropout of 0.1 probability is applied.

\subsubsection{Self-Attentive Encoder for Conditional Dropout Prediction}

Unlike DAS, which utilizes a transformer encoder-decoder structure, we choose a single layer of transformer encoder with causal masks for building an autoregressive classifier simply (see Fig. \ref{fig:model_diagrams} (c)). The transformer consists of 5 heads and 100 feedforward dimensions. Also, 0.2 dropout rate and ReLU activation are applied.

\subsection{Extra Information of Experiments}

\subsubsection{Used Features}

For knowledge tracing, as we denoted, we only used item id/skill id and correctness information to make our setup the same as other baselines. For statics 2011, concatenation of ‘Problem Name’ and ‘KC (Unique-step)’ is used as an item id. On the other hand, we use multiple features for dropout prediction for each task as different baselines did. For KDD15, module id, course id, event, object, category, source, open from, open to, start time, and lag time are used. The lag time is calculated from the start time, and we min-max normalize continuous features (open from, ‘open to, start time, and lag time) in advance.
Regarding the lag time, we set 0 as the minimum value and 604800000 ms as the maximum value. We clip that value to the maximum if any value exceeds the max limit. For XuetangX, we use module id, action, object, and lag time.
Similarly, the lag time is calculated from the start time and normalized. Meanwhile, we use item id, part id, correctness, and elapsed time information for conditional dropout predictions. Like lag time, we min-max normalize elapsed time in advance, setting the maximum as 180000 ms and 0 ms for the minimum value.

\subsubsection{Details of Training}
While we fix all batch sizes of baselines 128, we set 64 batch-size for pretraining but 128 batch-size for finetuning or multitask learning. In addition, we apply weight decay 1e-6 for all datasets to prevent overfitting each baseline. For ASSISTments 2009, 2015, we find that using more weight decay (1e-4) with contrastive methods during the pretrain stage can slightly help the model learn a better representation. After then, we set weight decay as 1e-6 on finetuning stage to be the same as other baselines.

\subsubsection{Setup of Data Augmentation}
On KT, for comparisons with \cite{lee2022cl4kt}, we change each data augmentation setup used in CL4KT. The code is from the original repository\footnote{https://github.com/UpstageAI/cl4kt}, but some setups are different, including data filtering, to be synced with other baseline setups. We select the best hyperparameters by changing of probabilities of each data augmentations ($\gamma_{mask}$, $\gamma_{crop}$, $\gamma_{replace}$, $\gamma_{permutation}$) within the range of \{0.3, 0.5, 0.7\} and reported the best results. For other settings, we follow the hyperparameter setups of the original repository. For ablation experiments about data augmentations, we use the best setups.

We use four data augmentations for comparisons on the DP and Cond-DP. Similar to the augmentation of CL4KT but an extended version to support multi-column features, we apply masking features, cropping some interactions, replacing items, and permuting the order of interactions.

We ran all our experiments on the NVIDIA RTX 2080 Ti.



\end{document}